\documentclass[12pt]{report}
\usepackage{amssymb}
\usepackage{amsmath}
\usepackage{graphicx}
\begin{document}
\voffset = -2.9cm
\hoffset = -1.3cm
\def\itm{\newline \makebox[8mm]{}}
\def\ls{\makebox[8mm]{}}
\def\fra#1#2{\frac{#1}{#2}}
\def\fr#1#2{#1/#2}
\def\frl#1#2{\mbox{\large $\frac{#1}{\rule[-0mm]{0mm}{3.15mm} #2}$}}
\def\frn#1#2{\mbox{\normalsize $\frac{#1}{\rule[-0mm]{0mm}{3.15mm} #2}$}}
\def\frm#1#2{\mbox{\normalsize $\frac{#1}{\rule[-0mm]{0mm}{2.85mm} #2}$}}
\def\frn#1#2{\mbox{\normalsize $\frac{#1}{\rule[-0mm]{0mm}{3.15mm} #2}$}}
\def\hs#1{\mbox{\hspace{#1}}}
\def\b{\begin{equation}}
\def\e{\end{equation}}
\def\arccot{\mbox{arccot}}
\vspace*{6mm}
\makebox[\textwidth][c]
{\large \bf{A spacetime with closed timelike geodesics everywhere}}
\vspace{4mm} \newline
\makebox[\textwidth][c]
{\normalsize \O yvind Gr\o n$^{*}$ and Steinar Johannesen$^*$}
\vspace{1mm} \newline
\makebox[\textwidth][c]
{\scriptsize $*$ Oslo University College, Department of Engineering,
P.O.Box 4 St.Olavs Plass, N-0130 Oslo, Norway}
%\vspace{-5mm} \newline
%\makebox[\textwidth][c]
%{\scriptsize $\dag$ Institute of Physics, University of Oslo,
%P.O. box 1048, N-0316 Oslo, Norway}
%
\vspace{6mm} \newline
{\bf \small Abstract} {\small
In the present article we find a new class of solutions of Einstein's
field equations. It describes stationary, cylindrically symmetric
spacetimes with closed timelike geodesics everywhere outside the
symmetry axis. These spacetimes contain a magnetic field parallel
to the axis, a perfect fluid with constant density and pressure,
and Lorentz invariant vacuum with energy density represented by
a negative cosmological constant.}
%
% 1
%
\vspace{10mm} \newline
{\bf 1. Introduction}
\vspace{3mm} \newline
Only three known solutions of Einstein's field equations representing
spacetimes in ${\bf R}^4$ with closed timelike geodesics have previously
been found. It was demonstrated by Steadman [1] that such curves exist
in the exterior van Stockum spacetime [2] representing spacetime inside
and outside a cylindrically symmetric rotating dust distribution. It has
also been shown by Bonnor and Steadman [3] that a spacetime with two
spinning particles, each one with a magnetic moment and mass equal
to their charge, permit special cases in which there exist closed
timelike geodesics. More recently Gr\o n and Johannesen [4] found
a class of solutions of Einstein's field equations with closed timelike
geodesics representing spacetime outside a spinning cosmic string
surrounded by a region of finite radial extension with vacuum energy
and a gas of non-spinning strings. In all of these spacetimes there
are closed timelike geodesics only at particular radii.
\itm In the present article we present a solution of Einstein's field
equations representing a cylindrically symmetric spacetime with closed
timelike geodesics everywhere outside the axis.
%
%
% 2
%
\vspace{10mm} \newline
{\bf 2. A spacetime with closed timelike geodesics everywhere}
\vspace{3mm} \newline
We shall here investigate a class of stationary cylindrically
symmetric spacetimes described by a line element of the form
\begin{equation} \label{e_1}
ds^2 = -(dt - 2 \hs{0.7mm} \omega \hs{0.7mm} a(r) \hs{0.7mm} d\phi)^2
+ b(r)^2 d\phi^2 + dr^2 + dz^2
\mbox{ ,}
\end{equation}
where $\omega$ is a constant, using units so that $c = 1$. The
coordinate time is shown on standard clocks at rest in the coordinate
system. The presence of the product term
$4 \hs{0.7mm} \omega \hs{0.7mm} a(r) \hs{0.7mm} dt d\phi$ in the
line element means that the coordinate clocks are not Einstein
synchronized. Furthermore we assume that there is no gap in the
coordinate time along a closed curve around the axis.
\itm With this form of the line element, Weyssenhoff's formula
for the vorticity gives [5]
\begin{equation} \label{e_2}
\Omega = \frl{a'}{b} \hs{0.8mm} \omega
\mbox{ ,}
\end{equation}
where $a'$ is the derivative of $a$ with respect to $r$. We shall consider
a space with constant vorticity, $\Omega = \omega$. Hence $b = a'$.
\itm In the present paper we shall search for solutions of Einstein's
field equations with closed timelike geodesics at all radii $r > 0$.
\itm Let us consider circular timelike curves in the plane
$z = \mbox{constant}$ with center on the z-axis. For such curves to
be closed in spacetime, the condition ${a'}^2 - 4 \omega^2 a^2 < 0$,
must be fullfilled [4]. Closed timelike geodesics of this type exist in
a region where in addition $a' (a'' - 4 \omega^2 a) = 0$.
We have two cases:
\newline
1. $a' = 0$ which is not permitted because it implies that
the determinant of the metric tensor vanishes.
\newline
2. $a' \ne 0$ and $a'' - 4 \omega^2 a = 0$ which gives that
\begin{equation} \label{e_8}
a = C_1 \cosh (2 \omega r) + C_2 \sinh (2 \omega r)
\mbox{ .}
\end{equation}
From the conditions above it follows that the constants $C_1$ and $C_2$
must fullfill $C_1 > C_2 \ge 0$, where we have assumed that $a > 0$,
which involves no loss of physical generality. Since these conditions
may be fullfilled everywhere outside the z-axis, we can conclude that
there are closed timelike geodesics of the type described above in the
entire spacetime outside the z-axis.
\itm Einstein's field equations now give for the mixed components of the
total energy momentum tensor
\begin{equation} \label{e_4}
\kappa T^{\mu}_{\hs{2.5mm} \nu} = \omega^2
\left[ \begin{array}{cccc}
\rule[-2mm]{0mm}{6.0mm}
1 & 0 & 0 & 0 \\
\rule[-2mm]{0mm}{6.0mm} 0 & 1 & 0 & 0 \\
\rule[-2mm]{0mm}{6.0mm} 0 & 0 & 1 & 0 \\
\rule[-2mm]{0mm}{6.0mm} 0 & 0 & 0 & 3
\end{array} \right] + \Lambda {\delta}^{\mu}_{\hs{2.5mm} \nu}
\mbox{ ,}
\end{equation}
where $\kappa$ is Einstein's gravitational constant, and
$\Lambda$ is the cosmological constant representing Lorentz
invariant vacuum energy (LIVE).
%
%
% 3
%
\vspace{10mm} \newline
{\bf 3. The physical contents of this spacetime}
\vspace{3mm} \newline
In addition to LIVE the spacetime is filled with a perfect fluid and a
cylindrically symmetric magnetic field along the z-axis.
\itm The energy momentum tensor of the perfect fluid has components
\begin{equation} \label{e_19}
T^{\mu}_{\hs{2.5mm} \nu} = (\rho + p) u^{\mu} u_{\nu}
+ p {\delta}^{\mu}_{\hs{2.5mm} \nu}
\mbox{ ,}
\end{equation}
where ${\delta}^{\mu}_{\hs{2.5mm} \nu}$ is the Kronecker symbol.
The fluid is at rest in the coordinate system. Then the contravariant
components of the fluid's $4$-velocity are $u^t = 1$, $u^i = 0$. The
corresponding non-zero covariant components are $u_t = -1$,
$u_{\phi} = 2 \omega a$. The non-zero mixed components of the
energy momentum tensor are
\begin{equation} \label{e_32}
T^{t}_{\hs{2.5mm} t} = - \rho \hs{1.0mm} , \hs{2.0mm}
T^{r}_{\hs{2.5mm} r} = T^{\phi}_{\hs{2.5mm} \phi} = T^{z}_{\hs{2.5mm} z}
= p \hs{1.0mm} , \hs{2.0mm}
T^{t}_{\hs{2.5mm} \phi} = 2 \omega a (\rho + p)
\mbox{ .}
\end{equation}
\itm The mixed components of the energy momentum tensor for the magnetic
field are
\begin{equation} \label{e_20}
T^{\mu}_{\hs{2.5mm} \nu} = F^{\mu \alpha} F_{\alpha \nu}
- \frl{1}{4} {\delta}^{\mu}_{\hs{2.5mm} \nu}
F_{\alpha \beta}  F^{\alpha \beta}
\mbox{ .}
\end{equation}
The non-vanishing covariant components of the field tensor are
\begin{equation} \label{e_33}
F_{r \phi} = - F_{\phi r} = B
\mbox{ ,}
\end{equation}
where $B$ is the magnetic field strength.
This gives the following non-vanishing mixed components of the energy
momentum tensor
\begin{equation} \label{e_34}
T^{t}_{\hs{2.5mm} t} = - T^{r}_{\hs{2.5mm} r}
= - T^{\phi}_{\hs{2.5mm} \phi} = T^{z}_{\hs{2.5mm} z} = \frl{B^2}{2 a'^2}
\hs{1.0mm} , \hs{2.0mm}
T^{t}_{\hs{2.5mm} \phi} = - \frl{2 \omega a}{a'^2} B^2
\mbox{ .}
\end{equation}
We then obtain the following independent field equations
\begin{equation} \label{e_21}
\omega^2 + \Lambda = \kappa \left( \frl{B^2}{2 a'^2} - \rho \right)
\mbox{ ,}
\end{equation}
\begin{equation} \label{e_22}
\omega^2 + \Lambda = \kappa \left( - \frl{B^2}{2 a'^2} + p \right)
\mbox{ ,}
\end{equation}
\begin{equation} \label{e_23}
3 \omega^2 + \Lambda = \kappa \left( \frl{B^2}{2 a'^2} + p \right)
\mbox{ .}
\end{equation}
Combining equations \eqref{e_21} and \eqref{e_22} we obtain
\begin{equation} \label{e_24}
B^2 = (\rho + p) a'^2
\end{equation}
in accordance with the field equation for $T^{t}_{\hs{2.5mm} \phi}$,
which hence follows from the other field equations.
Subtracting equation \eqref{e_22} from \eqref{e_23} and assuming that
the magnetic field points in the positive z-direction, we get
\begin{equation} \label{e_25}
B = \sqrt{\frl{2}{\kappa}} \hs{1.0mm} \omega a'
\mbox{ .}
\end{equation}
The last two equations give
\begin{equation} \label{e_26}
\kappa (\rho + p) = 2 \omega^2
\mbox{ ,}
\end{equation}
which is the equation of state of the perfect fluid.
Substituting equation \eqref{e_25} into equation \eqref{e_21} leads
to
\begin{equation} \label{e_27}
\kappa \rho = - \Lambda = - \kappa {\rho}_v
\mbox{ ,}
\end{equation}
where ${\rho}_v$ is the density of the vacuum energy.
Equations \eqref{e_26} and \eqref{e_27} show that the density and
pressure of the perfect fluid are constant, and that the cosmological
constant must be negative in order that the density of the perfect
fluid shall be positive.
The total energy density of the perfect fluid and the vacuum energy
vanishes. The energy density in the spacetime comes from the magnetic
field.
%
%
% 4
%
%\vspace{10mm} \newline
\newpage
{\bf 4. A simple special case}
\vspace{3mm} \newline
%
%
% 4.1
%
{\bf 4.1. Kinematics}
\vspace{3mm} \newline
As a simple illustrating example we will consider a solution of the
field equations with $C_1 = 1$ and $C_2 = 0$ in equation \eqref{e_8},
giving
\begin{equation} \label{e_38}
a = \cosh (2 \omega r)
\mbox{ .}
\end{equation}
%.
The line element then takes the form
\begin{equation} \label{e_28}
ds^2 = -dt^2 + 4 \hs{0.7mm} \omega \hs{0.7mm} \cosh (2 \omega r)
\hs{0.7mm} d\phi dt - 4 \omega^2 d\phi^2 + dr^2 + dz^2
\mbox{ ,}
\end{equation}
In spite of the minus sign in front of $d\phi^2$ the signature is correct.
This is seen from the form \eqref{e_1} of the line element corresponding
to the orthogonal basis $({\bf e}_t,
{\bf e}_{\phi} + 2 \omega a(r) {\bf e}_t, {\bf e}_r,{\bf e}_z)$, since
${\bf e}_t \cdot {\bf e}_t < 0$,
$({\bf e}_{\phi} + 2 \omega a(r) {\bf e}_t) \cdot
({\bf e}_{\phi} + 2 \omega a(r) {\bf e}_t) = b(r)^2 > 0$,
${\bf e}_r \cdot {\bf e}_r > 0$ and
${\bf e}_z \cdot {\bf e}_z > 0$.
\itm The 3-space defined by the simultaneity $t = \mbox{constant}$ is
given by the spatial line element
\begin{equation} \label{e_29}
dl^2 = - 4 \omega^2 d\phi^2 + dr^2 + dz^2
\mbox{ .}
\end{equation}
This shows that the vector ${\bf e}_{\phi}$, which is a tangent vector
in this space, has ${\bf e}_{\phi} \cdot {\bf e}_{\phi}
= - 4 \omega^2 < 0$. Hence this vector is timelike.
\itm In the present spacetime the equation of circular null curves
in the plane $z = \mbox{constant}$ with center on the z-axis is
\begin{equation} \label{e_30}
4 \omega^2 \left( \frl{d \phi}{dt} \right)^2
- 4 \omega \cosh (2 \omega r) \frl{d \phi}{dt} + 1 = 0
\mbox{ .}
\end{equation}
The physical velocities of light moving in opposite
$\phi$-directions are
\begin{equation} \label{e_31}
v_{\pm} = \sqrt{|g_{\phi \phi}|} \left( \frl{d \phi}{dt} \right)_{\pm} =
2 \omega \left( \frl{d \phi}{dt} \right)_{\pm} = e^{\pm 2 \omega r}
\mbox{ .}
\end{equation}
The speed of light is diffent from $1$ since the coordinate clocks are
not Einstein synchronized. These expressions give the intersections of
light cones with the plane in the tangent space spanned by
${\bf e}_{\phi}$ and ${\bf e}_{t}$ as shown in Figure 1. The reason
for the direction of the light cone is the following. We have considered
light signals submitted in the positive and the negative $\phi$-direction.
Their angular velocities are denoted by plus and minus in equation
$\eqref{e_31}$. One signal has $d \phi > 0$ and the other $d \phi < 0$,
but the angular velocities of both have the same sign. Hence the
coordinate time interval $dt > 0$ for the first signal, and $dt < 0$
for the second one. Also $| (d \phi / dt)_{+}| > |(d \phi / dt)_{-}|$.
\vspace{5mm} \newline
%\begin{picture}(50,0)(-96,-152)
%\begin{picture}(50,0)(-96,-167)
\begin{picture}(50,175)(-96,-172)
%\input{fig4}
%
% Figure 1
%
%
\qbezier(154.8549, -40.4686)(154.1718, -39.5945)(150.8827, -41.1492)
\qbezier(150.8827, -41.1492)(147.5936, -42.7039)(142.0548, -46.5189)
\qbezier(142.0548, -46.5189)(136.5160, -50.3340)(129.3278, -55.9959)
\qbezier(129.3278, -55.9959)(122.1395, -61.6578)(114.0808, -68.5530)
\qbezier(114.0808, -68.5530)(106.0221, -75.4482)( 97.9662, -82.8295)
\qbezier( 97.9662, -82.8295)( 89.9103, -90.2109)( 82.7301, -97.2785)
\qbezier( 82.7301, -97.2785)( 75.5500, -104.3461)( 70.0237, -110.3340)
\qbezier( 70.0237, -110.3340)( 64.4974, -116.3219)( 61.2238, -120.5814)
\qbezier( 61.2238, -120.5814)( 57.9503, -124.8408)( 57.2842, -126.9101)
\qbezier( 57.2842, -126.9101)( 56.6181, -128.9794)( 58.6316, -128.6344)
\qbezier( 58.6316, -128.6344)( 60.6452, -128.2893)( 65.1202, -125.5673)
\qbezier( 65.1202, -125.5673)( 69.5952, -122.8453)( 76.0467, -118.0414)
\qbezier( 76.0467, -118.0414)( 82.4981, -113.2374)( 90.2270, -106.8720)
\qbezier( 90.2270, -106.8720)( 97.9559, -100.5066)(106.1247, -93.2696)
\qbezier(106.1247, -93.2696)(114.2934, -86.0327)(122.0168, -78.7083)
\qbezier(122.0168, -78.7083)(129.7401, -71.3839)(136.1812, -64.7659)
\qbezier(136.1812, -64.7659)(142.6222, -58.1479)(147.0830, -52.9534)
\qbezier(147.0830, -52.9534)(151.5438, -47.7588)(153.5408, -44.5507)
\qbezier(153.5408, -44.5507)(155.5379, -41.3426)(154.8549, -40.4686)
\qbezier(127.3585, -111.6574)(141.1067, -76.0630)(154.8549, -40.4686)
\qbezier(127.3585, -111.6574)( 92.3255, -120.0145)( 57.2926, -128.3717)
\put( 50.9434, -111.6574){\line(1, 0){ 33.3267}}
\put(127.3585, -111.6574){\vector(1, 0){ 76.4151}}
\put( 88.1035, -111.6574){\line(1, 0){  3.8335}}
\put( 95.7705, -111.6574){\line(1, 0){  3.8335}}
\put(103.4375, -111.6574){\line(1, 0){  3.8335}}
\put(111.1045, -111.6574){\line(1, 0){  3.8335}}
\put(118.7715, -111.6574){\line(1, 0){  3.8335}}
\put(127.3585, -166.3586){\line(0, 1){ 54.7012}}
\put(127.3585, -73.6231){\vector(0, 1){ 57.6928}}
\put(127.3585, -107.5411){\line(0, 1){  4.1163}}
\put(127.3585, -99.3086){\line(0, 1){  4.1163}}
\put(127.3585, -91.0761){\line(0, 1){  4.1163}}
\put(127.3585, -82.8436){\line(0, 1){  4.1163}}
\put(127.3585, -111.6574){\vector(-2, -1){ 50.9434}}
\put(127.3585, -111.6574){\line(2, 1){ 50.9434}}
\put( 70.0472, -143.1106){\makebox(0,0)[]{\normalsize{$x$}}}
\put(213.9623, -115.7600){\makebox(0,0)[]{\normalsize{$y$}}}
\put(119.7170, -14.5628){\makebox(0,0)[]{\normalsize{$t$}}}
\end{picture}
%\vspace{10mm} \newline
\vspace{3mm} \newline
\hspace*{3mm} \parbox[t]{150mm}{\footnotesize \sf {\bf Figure 1.}
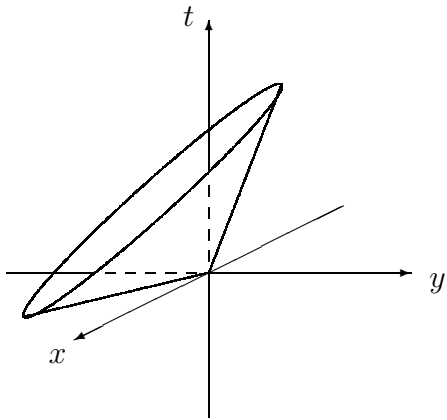
Light cones in the spacetime with line element $\eqref{e_28}$ where
the x-, y- and t-axes correspond to the basis vectors ${\bf e}_{r}$,
${\bf e}_{\phi}$ and ${\bf e}_{t}$.}
\vspace{5mm} \newline
\itm This figure shows that the tangent vector to a curve in the $\phi$
direction with $t = \mbox{constant}$ is inside the light cone at
each point. Hence such a curve is timelike.
\itm The light signal travelling in the negative
$\phi$-direction actually travels backwards in time, and hence can
be used to warn people living in previous times about possible
catastrophic events in their future. This also involves the
possibility of causal paradoxes, and may demand some sort of
chronology protection [6].
\itm Let us calculate how far backwards in time light can come by
travelling one time around the axis. Integrating equation \eqref{e_31}
we find
\begin{equation} \label{e_39}
\Delta t_1 = - 4 \pi \omega e^{- 2 \omega r_1}
\end{equation}
for light travelling in the negative $\phi$-direction. Hence by
travelling an arbitrary number of times around the axis the light
may arrive arbitrarily far backwards in time.
\itm We now consider light moving in both the $\phi$- and $r$-directions.
Then the 4-velocity identity for the light takes the form
\begin{equation} \label{e_35}
-\dot{t}^2 + 4 \hs{0.7mm} \omega \hs{0.7mm} \cosh (2 \omega r)
\hs{0.7mm} \dot{\phi} \hs{0.7mm} \dot{t} - 4 \omega^2 \dot{\phi}^2
+ \dot{r}^2 = 0
\mbox{ ,}
\end{equation}
where the dot denotes differentiation with respect to an invariant
parameter. This equation shows that there exist null curves with
$\dot{t} = 0$ given by
\begin{equation} \label{e_36}
4 \omega^2 \dot{\phi}^2 = \dot{r}^2
\mbox{ .}
\end{equation}
Integrating this equation, we obtain
\begin{equation} \label{e_37}
r = r_0 \pm 2 \omega \phi
\end{equation}
describing Archimedean spirals. Light signals moving along these curves
travel infinitely fast, i.e. they arrive at the same point of time $t$
as they are emitted.
%
%
%
% 4.2
%
\newpage
%\vspace{10mm} \newline
{\bf 4.2. Geodesics}
\vspace{3mm} \newline
The null curves that we have considered are not in general geodesics.
We shall now consider timelike and null geodesics. The Lagrangian
function of a materal particle or a photon moving in a plane
$z = \mbox{constant}$ in a spacetime described by a line element
\eqref{e_28} is
\begin{equation} \label{e_42}
L = - \frl{1}{2} \dot{t}^2
+ 2 \hs{0.7mm} \omega \hs{0.7mm} \cosh (2 \omega r)
\hs{0.7mm} \dot{\phi} \hs{0.7mm} \dot{t}
- 2 \omega^2 \dot{\phi}^2 + \frl{1}{2} \dot{r}^2
\mbox{ .}
\end{equation}
Then the $\phi$-component of the geodesic equation is $\ddot{\phi} = 0$,
implying that $\dot{\phi} = \mbox{constant}$ along the path. The radial
component of the geodesic equation is
\begin{equation} \label{e_43}
\ddot{r} = 2 \omega^2 \sinh (2 \omega r) \hs{0.7mm}
\dot{\phi} \hs{0.7mm} \dot{t}
\mbox{ .}
\end{equation}
Hence circular geodesics with $\ddot{r} = 0$ must have $\dot{t} = 0$,
and are therefore closed in spacetime. This is the case for both
timelike and null geodesics. Thus the null curves going backwards in time,
which were considered above, are not geodesics. These photons must
move for instance in fibre optic cables.
\itm Considering a material particle moving along a circle in the
$z = \mbox{constant}$ plane with center on the z-axis, the 4-velocity
identity takes the form
\begin{equation} \label{e_40}
-\dot{t}^2 + 4 \hs{0.7mm} \omega \hs{0.7mm} \cosh (2 \omega r)
\hs{0.7mm} \dot{\phi} \hs{0.7mm} \dot{t} - 4 \omega^2 \dot{\phi}^2 = -1
\mbox{ .}
\end{equation}
If this circle is closed in spacetime, $\dot{t} = 0$, the particle
must have
\begin{equation} \label{e_41}
(\dot{\phi})_{\pm} = \pm \frl{1}{2 \omega}
\mbox{ .}
\end{equation}
It can be shown that the geodesic equation is fullfilled for particles
moving in this way. Hence particles following closed timelike
geodesics in spacetime move with constant angular velocity.
\itm Since $\phi$ is a cyclic coordinate
$p_{\phi} = l \dot{\phi} + k \dot{t} = \mbox{constant}$, where
$k = 2 \omega a$ and $l = a'^2 - 4 \omega^2  a^2$.
For $\dot{t} = 0$ this gives $l \dot{\phi} = \mbox{constant}$.
For circular motion the 4-velocity identity of a material particle
reduces to $- \dot{t}^2 + 2 k \dot{t} \dot{\phi} + l \dot{\phi}^2 = -1$.
In order that the circle shall be closed in spacetime, the condition
$\dot{t} = 0$ must be fullfilled, which gives $l \dot{\phi}^2 = -1$.
Hence $\dot{\phi} = \mbox{constant}$ along the closed timelike geodesic.
In general this constant has an $r$-dependent value, permitting $l$ to
be a function of $r$ as in [4].
\itm The fact that the angular velocity is constant is a consequence of
the axial symmetry, and is valid for arbitrary functions $a(r)$.
%
%
% 5
%
\vspace{10mm} \newline
{\bf 5. Conclusion}
\vspace{3mm} \newline
We have constructed solutions of Einstein's field equations in a
cylindrically symmetric space with closed timelike geodesics in the
entire space outside the symmetry axis.
There exist circular geodesics in the plane $z = \mbox{constant}$
with center on the z-axis so that a particle moving freely along one
of these geodesics arrives at the same event as it departed from.
In addition there exist such timelike and null circular worldlines of
material particles or photons going backwards in time. However, these
worldlines are not geodesics.
\newpage
%\vspace{5mm} \newline
{\bf References}
%\vspace{3mm} \newline
%
\begin{enumerate}
\item B.R.Steadman, {\it Causality Violation on van Stockum Geodesics},
Gen.Rel.Grav. {\bf 35}, 1721 (2003).
\item W.J. van Stockum, Pros.R.Soc.Edinb. {\bf 57}, 135 (1937).
\item W.B.Bonnor and B.R.Steadman, {\it Exact solutions of the
Einstein Maxwell equations with closed timelike curves.},
Gen.Rel.Grav. {\bf 37}, 1833 (2005).
\item \O.Gr\o n and S.Johannesen, {\it Closed timelike geodesics in a gas
of cosmic strings}, New Journal of Physics, {\bf 10}, 103025 (2008).
\item M.J.Weyssenhoff, \textit{Metrisches Feld und Gravitationssfeld},
Bull. Acad. Pol. Sci. Lett. A 252 (1937).
\item S.W.Hawking, \textit{The chronology protection conjecture},
Phys.Rev. D46 603 - 611 (1992)
\end{enumerate}
\end{document}